\documentclass[aps,pra,reprint,amsmath,amssymb,superscriptaddress]{revtex4-1} 
\usepackage{graphicx}
\usepackage{dcolumn}
\usepackage{xcolor}
\usepackage{bm}
\usepackage{float}
\usepackage{mathrsfs}
\usepackage{times}
\usepackage{hyperref}
\usepackage{physics}
\usepackage{bbold}

\begin{document}	
	
\newcommand{\titleinfo}{Quantum simulation of hadronic states with Rydberg-dressed atoms}
	
\title{\titleinfo}
\author{Zihan Wang}\email{zihan.wang18@imperial.ac.uk}
\affiliation{Blackett Laboratory, Imperial College London, London SW7 2AZ, United Kingdom}

\author{Feiyang Wang}
\affiliation{Blackett Laboratory, Imperial College London, London SW7 2AZ, United Kingdom}

\author{Joseph Vovrosh}
\affiliation{Blackett Laboratory, Imperial College London, London SW7 2AZ, United Kingdom}
\affiliation{PASQAL SAS, 2 av. Augustin Fresnel, 91120 Palaiseau, France}

\author{Johannes Knolle}
\affiliation{Blackett Laboratory, Imperial College London, London SW7 2AZ, United Kingdom}
\affiliation{Munich Center for Quantum Science and Technology (MCQST), Schellingstr. 4, D-80799 M\"{u}nchen, Germany}
\affiliation{Department of Physics TQM, Technical University of Munich, 85748 Garching, Germany}
	
\author{Florian Mintert}
\affiliation{Blackett Laboratory, Imperial College London, London SW7 2AZ, United Kingdom}
\affiliation{Helmholtz-Zentrum Dresden-Rossendorf, Bautzner Landstra{\ss}e 400, 01328 Dresden, Germany}
	
\author{Rick Mukherjee}\email{rick.mukherjee@physnet.uni-hamburg.de}
\affiliation{Blackett Laboratory, Imperial College London, London SW7 2AZ, United Kingdom}
\affiliation{Center for Optical Quantum Technologies, Department of Physics, University of Hamburg, Luruper Chaussee 149, 22761 Hamburg, Germany}

\begin{abstract}
The phenomenon of confinement is well known in high-energy physics and can also be realized for low-energy domain-wall excitations in one-dimensional quantum spin chains. A bound state consisting of two domain-walls can behave like a meson, and in a recent work of Vovrosh \textit{et al.} [PRX Quantum 3, 040309 (2022)] , it was demonstrated that a pair of mesons could dynamically form a meta-stable confinement-induced bound state (consisting of four domain-walls) akin to a hadronic state. However, the protocol discussed in Vovrosh \textit{et al.} [PRX Quantum 3, 040309 (2022)] involving the use of interactions with characteristically non-monotonic distance dependence is not easy to come by in nature, thus, posing a challenge for its experimental realization. In this regard, Rydberg atoms can provide the required platform for simulating confinement-related physics. We exploit the flexibility offered by interacting Rydberg-dressed atoms to engineering modified spin-spin interactions for the one-dimensional transverse field Ising model. Our numerical simulations show how Rydberg-dressed interactions can give rise to a variety of effective potentials that are suitable for hadron formation, which opens the possibility of simulating confinement physics with Rydberg platforms as a viable alternative to current trapped-ion experiments.
\end{abstract}

\maketitle

\textit{Introduction.\textemdash} Confinement is one of the cornerstones of high-energy physics. In quantum chromodynamics (QCD), the asymptotic freedom of the strong interaction implies that the interaction energy between a pair of elementary particles such as quarks grows with increasing separation \cite{greensite2011introduction}. As a result of this, quarks remain confined within a certain distance forming composite objects such as hadrons.  Since high-energy experiments are extremely costly and resource intensive, there is increasing interest in quantum simulations that can guide experimental activities \cite{Carmen, Surace, Yang_gauge, Zhao_gauge}. 

\begin{figure}[t!]
	\centering
	\includegraphics[width = \columnwidth]{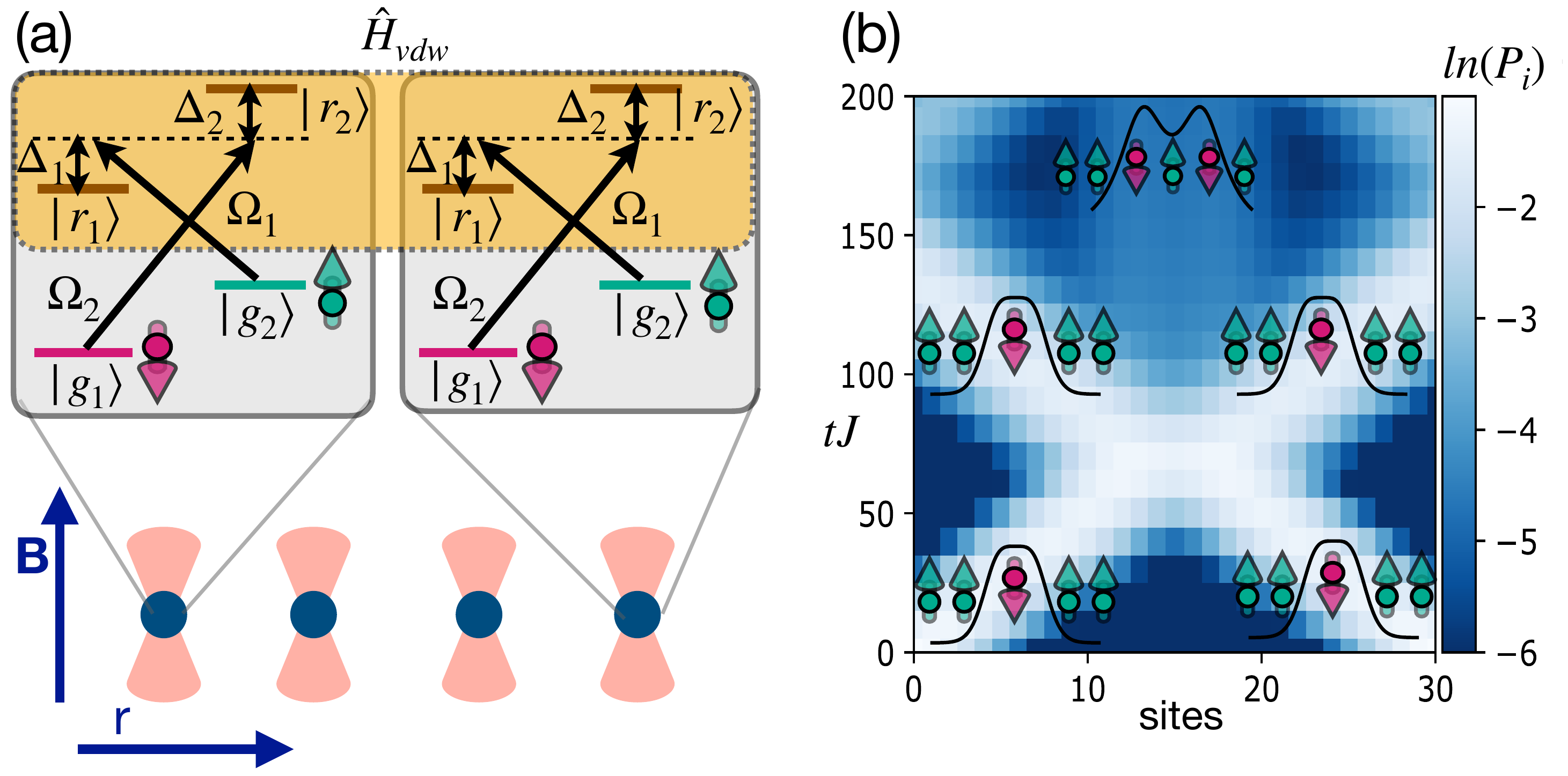}
	\caption{(a) Setup involving an array of trapped Rydberg-dressed atoms simulating the spin-1/2 system: Each atom has a four-level structure with two ground states ($\ket{g_1},\ket{g_2}$) coupled to Rydberg states ($\ket{r_1}, \ket{r_2}$) with a pair of excitation lasers such that $\ket{g_i}$ only couples to $\ket{r_{j\neq i}}$ with Rabi frequency $\Omega_j$ and detuning $\Delta_j$. Atoms interact via the Rydberg-Rydberg interaction, which is highlighted in the top orange box denoted by $\hat{H}_{\mathrm{vdw}}$. (b) Dynamical formation of a tetraquark during a collision event between mesons: Numerical results showing dynamics of two $1$-meson wavepackets (meson of width 1 site as shown schematically) where the majority of the wavepackets are reflected elastically, but with a small and finite probability to form a tetraquark at the center. Here $P_i$ measures the probability that a domain-wall is located at site $i$. Parameters used for the numerical simulation are discussed in Sec. 5 of the Appendix.}
	\label{Fig1_Setup}
\end{figure}

Interestingly, various aspects of confinement physics are realized in condensed matter systems, in particular, one-dimensional quantum spin chains \cite{mccoy1978two, Delfino_1996, kormos2017real, liu2019confined}. In such models, the initial state acts as a source of quasiparticle excitations that are well described by doman-walls which are regions of up (down) spins followed by a region of down (up) spins as shown schematically in Fig.~\ref{Fig1_Setup}(b). It has been shown that bound states of domain-walls which behave like mesons (two-domain-wall state) can appear in the transverse field Ising model (TFIM) with a longitudinal field \cite{kormos2017real} or with long-range interactions \cite{liu2019confined}. Although confinement in one-dimensional spin chains is much simpler than confinement physics in higher dimensions studied in QCD, it is nevertheless a noteworthy step in quantum simulation.

The formation of meson bound states in one dimensional quantum spin models has interesting consequences for non-equilibrium physics \cite{kormos2017real}, the propagation of energy \cite{mazza2019suppression}, and quantum information \cite{vovrosh2021confinement, PhysRevE.104.035309}. Although scattering processes between mesons  are predominantly elastic, the formation of a stable multi-meson bound state such as a tetra-quark (four-domain wall state) requires deep inelastic collisions \cite{milsted2020collisions,karpov2020spatiotemporal,surace2021scattering}. One way to enhance the overlap of the initial asymptotic scattering free meson states with the hadronic bound states is by quenching the kinetic energy of the individual mesons. This approach requires a high degree of precision and control in the temporal variation of the external field. An alternative approach to obtain hadronic bound states is to fuse a pair of mesons by inducing non-monotonic meson-meson interaction \cite{vovrosh2022dynamical}. However, realizing non-monotonic spin interactions in conventional simulators, such as trapped ions or cold atoms, is challenging in general.

In this article, we detail the microscopic implementation of dynamical hadron formation on a Rydberg quantum simulator and identify parameters suitable for the experimental realization of the protocol discussed in Ref.~\cite{vovrosh2022dynamical}. Rydberg dressing \cite{Che2015, Zeiher2016,Balewski_2014, Young2018, Madjarov2020, Henkel2010, Tyler2013, Petrosyan2014, Glaetzle2015, Li2018, Wu2021, Gil, Bijnen, charge, Kele2020} is a well-known and useful concept that allows for a small admixture of Rydberg state to the ground state atoms, which in turn provides unique flexibility to manipulate the strength and shape of the interactions using parameters that are easily tunable in experiments. The goal is to exploit these highly tunable anisotropic Rydberg-dressed interactions to provide the necessary non-monotonic potential that will facilitate the formation of the hadron during the mesonic dynamics. 
\begin{figure}
	\centering
	\includegraphics[width = 0.83\columnwidth]{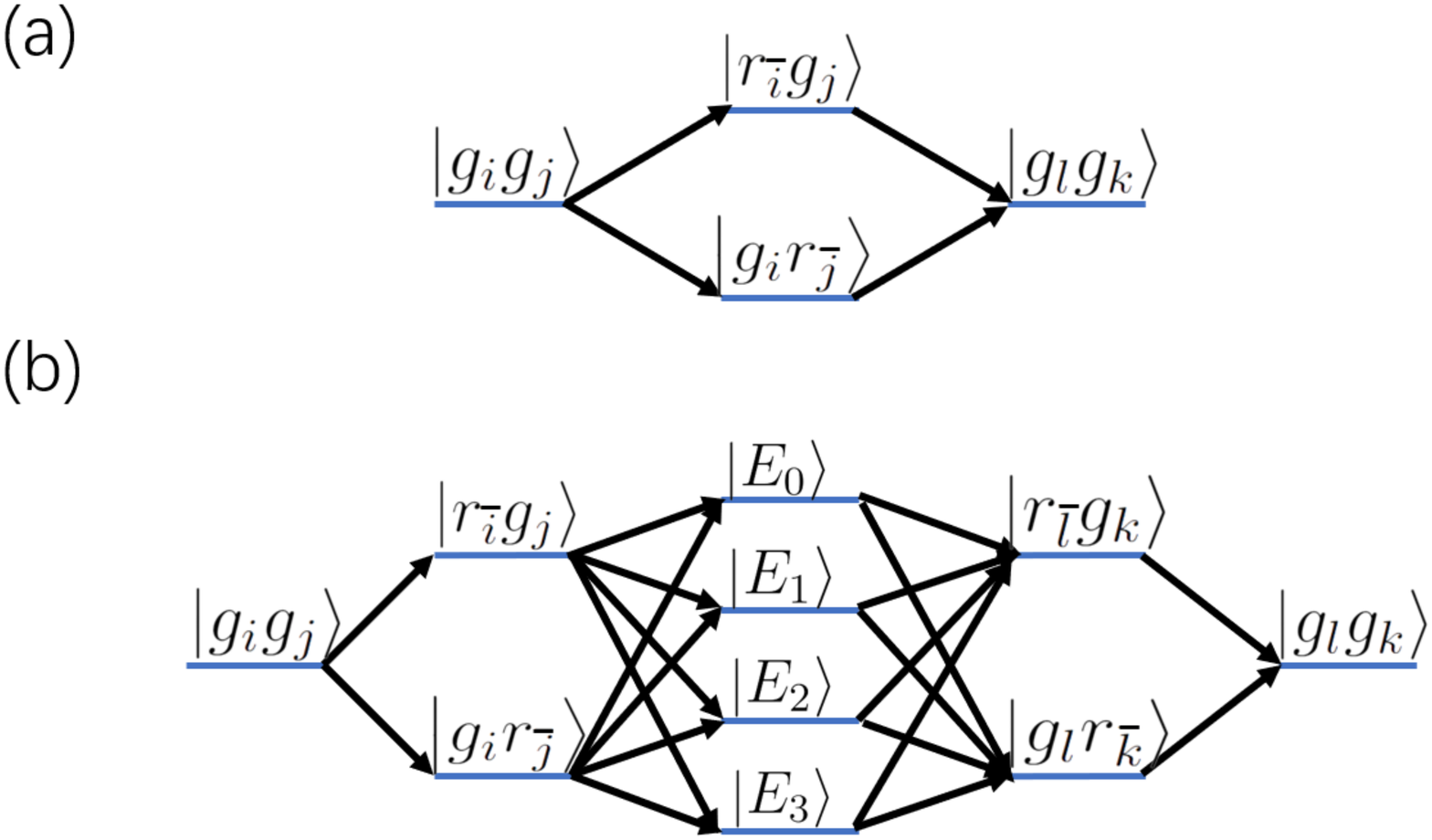}
	\caption{Schematic diagrams outlining the different possible channels during adiabatic elimination of the different Rydberg states belonging to the same Zeeman manifold in the weak coupling limit: Second-order processes contain only singly excited states as intermediate channels as shown in (a), while fourth-order processes also include eigenstates $\ket{E_{i=0,1,2,3}}$  which are the result of the Rydberg interaction induced Zeeman mixing when both atoms are in Rydberg states. $\{i,j,k,l\} \in \{1,2\}$ with bar representing the compliment of value.}
	\label{leveldg}
\end{figure} 

\textit{Setup.\textemdash} Consider a one-dimensional array of trapped atoms with each atom possessing a pair of  ground states $\ket{g_1}$ and $\ket{g_2}$ that are weakly coupled to Rydberg states $\ket{r_2}$ and $\ket{r_1}$ respectively using excitation lasers that are characterized by Rabi frequencies $\Omega_{i=1,2}$ and detunings $\Delta_{i=1,2}$ as shown in Fig.~\ref{Fig1_Setup}(a). A magnetic field along the direction of the laser is applied shown as $B$ in Fig.~\ref{Fig1_Setup}(a) which defines the quantization axis and  by choosing (left/right) circularly polarized excitation lasers, one obtains the level scheme described in Fig.~\ref{Fig1_Setup}(a). The indices $\{i,j,l,k\}$ label the internal levels for the individual atoms and not the position of the atoms. The ground state energy $E_{g}$ is set to zero without loss of generality. A pair of atoms separated by distance $\mathbf{r}$ can interact only when they are both excited to their Rydberg states. Focusing on this two-atom picture, the Hamiltonian that describes the laser excitation process along with the atom-atom interaction is given as
\begin{equation} \label{hamiltonians}
	\begin{aligned}
		&\hat{H} = \hat{H}_{0}(\mathbf{r}) + \hat{H}' ,\\
		&\hat{H}_{0}(\mathbf{r}) = \sum_{i=1}^{2} -\Delta_{i}\ket{r_i}\bra{r_i} + \hat{H}_{\mathrm{vdw}}(\mathbf{r}) ,\\
		&\hat{H}' = \sum_{i,j=1;i\ne j}^{2} \left(\frac{\Omega_{j}}{2}\ket{g_i}\bra{r_j} + h.c. \right).
	\end{aligned}
\end{equation}
Here $\hat{H}_{\mathrm{vdw}}(\mathbf{r})$ is the Hamiltonian that describes the van der Waals interaction between two Rydberg atoms assuming that the Rydberg states are far away from any F\"{o}rster resonance and is given as 
\begin{equation}\label{vdweq}
	\hat{H}_{\mathrm{vdw}}(\mathbf{r}) = \sum_{\eta}  \frac{C_6(\eta)}{r^6} \mathcal{\hat{D}}_{\eta}(\theta,\phi) .
\end{equation}
The van der Waals interactions have two effects: a shift in Rydberg levels and a mixing of different Rydberg states within the same Zeeman manifold \cite{Walker}. Depending on the choice of the Rydberg states ($\ket{r_{i=1,2}}$), there are different channels of intermediate states that couple to $\ket{r_{i=1,2}}$ which are denoted by $\eta$. The overall strength of the Rydberg interaction is determined by summing over all the relevant channels as defined in Eq.~\ref{vdweq}. For each channel, there is a van der Waals coefficient $C_6(\eta)$ which depends on the radial component of the dipole matrix element, while $\mathcal{\hat{D}}_{\eta}(\theta,\phi)$ is the operator that projects onto the subspace of magnetic quantum numbers allowed by the selection rules. Thus $\mathcal{\hat{D}}_{\eta}(\theta,\phi)$ has the information about the anisotropy in the Rydberg interactions. Explicit expressions for $C_6(\eta)$ and $\mathcal{\hat{D}}_{\eta}(\theta,\phi)$ are provided in Sec.~1 of the Appendix.

\textit{Rydberg-dressing.\textemdash} Although bare ground state atoms do not interact, effective interactions between ground state atoms can be induced as a result of Rydberg-dressing \cite{Glaetzle2015}. Rydberg dressing is a consequence of weak optical coupling of the ground states to their corresponding Rydberg states, which is satisfied when $\Delta_{1,2} \gg\Omega_{1,2}$ \cite{Johnson2010,Honer2010,Balewski_2014, Mukherjee2}. In this limit, perturbation theory is used to determine $\tilde{V}^{ij}_{lk}(\mathbf{r})$ which denote the effective interactions between a pair of Rydberg-dressed atoms with ground states $\ket{\tilde{g}_{i,j=1,2}}$ and $\ket{\tilde{g}_{l,k=1,2}}$ that are separated by a distance vector $\mathbf{r}$ \cite{Glaetzle2015}. The final expressions $\tilde{V}^{ij}_{lk}$ depend on the bare Rydberg interactions as well as the laser parameters $(\Omega,\Delta)$ and have the following form,
\begin{equation}
H_{\text{eff}} = \sum_{\substack{i,j \\ l,k}} 	\tilde{V}^{ij}_{lk}(\mathbf{r}) \ket{\tilde{g}_{i}\tilde{g}_{j}}\bra{\tilde{g}_{l}\tilde{g}_{k}} ,
\end{equation}	
where
\begin{align}
	\tilde{V}^{ij}_{lk}(\mathbf{r}) &= U^{ij(2)}_{lk}(\mathbf{r}) + U^{ij(4)}_{lk}(\mathbf{r}), \\
	U^{ij(2)}_{lk}(\mathbf{r}) &= \sum_{a}  \frac{\bra{g_ig_j}\hat{H}'\ket{a}\bra{a}\hat{H}'\ket{g_lg_k}}{E_{g}-E_{a}}, \label{u2} \\	
	U^{ij(4)}_{lk}(\mathbf{r}) &=\sum_{a b c} \frac{\langle g_ig_j|\hat{H}'| a\rangle\langle a|\hat{H}'| b\rangle\langle b|\hat{H}'| c\rangle\langle c|\hat{H}'| g_lg_k\rangle}{\left(E_{g}-E_{a}\right)\left(E_{g}-E_{b}\right)\left(E_{g}-E_{c}\right)} \nonumber \\
	&-U^{ij(2)}_{lk} \sum_{a} \frac{\bra{g_ig_j}\hat{H}'\ket{a}\bra{a}\hat{H}'\ket{g_lg_k}}{\left(E_{g}-E_{a}\right)^{2}}.\label{u4}
\end{align}
The effective interaction between the dressed ground states has a non-zero contribution from the even order terms in the perturbation theory indicated by $U^{ij(2,4)}_{lk}$. Higher than fourth-order processes are negligible for our purposes. The different channels that contribute to the second and fourth-order processes have been illustrated in Fig.~\ref{leveldg}(a,b). The two-atom states with single Rydberg excitation are labeled as $\ket{a}, \ket{c} = \{ \ket{r_1g_2},  \ket{r_2g_1},  \ket{g_1r_2}, \ket{g_2r_1} \}$ with energy $E_{a,c}$ and $\ket{b}=\ket{E_{i=0,1,2,3}}$ correspond to both atoms excited to Rydberg states with energy $E_{b}=E_{i=0,1,2,3}$. The energy eigenstates $\ket{E_{i=0,1,2,3}}$ are obtained by diagonalizing $\hat{H}_0$, which already take into account the Zeeman mixing of relevant states. The strength of Rydberg-dressed interactions is controlled by laser parameters, which are encapsulated in Eq.~\ref{u2}-\ref{u4}. The most generalized form for $\tilde{V}$ and $U$ for arbitrary $(\theta, \phi)$ is complex but can be evaluated numerically. Nevertheless, for a specific case of $(\theta=0, \phi=0)$, one obtains relatively simple analytical expressions, which are provided in Sec.~2 of the Appendix. 

\textit{Mapping to spin models.\textemdash} The two-atom interaction picture in the previous section is now generalized to the spin model. The Rydberg-dressed ground states $\ket{\tilde{g}_1}=\ket{\downarrow}$ and $\ket{\tilde{g}_2}=\ket{\uparrow}$ have small admixtures of Rydberg states into the bare ground states and are used to define the spin-1/2 system. Using the standard definitions of the Pauli spin operators for $\hat{\sigma}^z = 1/2(\ket{\uparrow}\bra{\uparrow} - \ket{\downarrow}\bra{\downarrow})$ and $\hat{\sigma}^x = 1/2(\ket{\uparrow}\bra{\downarrow} + \ket{\downarrow}\bra{\uparrow})$, the one-dimensional TFIM for $N$ spins is written as
\begin{align}\label{Ising}
	\hat{H} =& \sum_{p<q} \left[ J_{z}(\mathbf{r}_{pq}) \hat{\sigma}^z_p \hat{\sigma}^z_{q} + h_{x}(\mathbf{r}_{pq}) \hat{\sigma}^x_q(\mathbf{r}_{pq}) + h_z(\mathbf{r}_{pq}) \hat{\sigma}^z_q \right],\\			
	J_z(\mathbf{r}) =& \frac{1}{4}\left(\tilde{V}^{11}_{11}(\mathbf{r}) - \tilde{V}^{12}_{12}(\mathbf{r}) - \tilde{V}^{21}_{21}(\mathbf{r}) + \tilde{V}^{22}_{22}(\mathbf{r})\right) ,\\
	h_x(\mathbf{r}) =& \frac{1}{2} \left( \tilde{V}^{12}_{11}(\mathbf{r}) + \tilde{V}^{21}_{11}(\mathbf{r}) + \tilde{V}^{11}_{12}(\mathbf{r}) + \tilde{V}^{11}_{21}(\mathbf{r})
	 \right. \nonumber \\
	&\left. + \tilde{V}^{22}_{12}(\mathbf{r})  + \tilde{V}^{22}_{21}(\mathbf{r}) +\tilde{V}^{12}_{22}(\mathbf{r})  + \tilde{V}^{21}_{22}(\mathbf{r})     \right),  \\
	h_z(\mathbf{r}) =& \frac{1}{4} \left(\tilde{V}^{22}_{22}(\mathbf{r})- \tilde{V}^{11}_{11}(\mathbf{r})\right) ,
\end{align}
with $p,q$ labelling the lattice sites with position vectors $\mathbf{r}_{p},\mathbf{r}_{q}$ respectively and $\mathbf{r}_{pq} = \mathbf{r}_{p}- \mathbf{r}_{q}$ is the distance vector between the sites. The $h_{x/z}$ are the transverse/longitudinal field strengths, respectively, which in the Rydberg-dressed picture also depend on the distance vector as shown for $h_x$. The term $J_{z}(\mathbf{r})=J_z(r,\theta,\phi)$ determines the strength of the Ising interaction between spins with distance $r$ away from each other and depends on the choice of the principal quantum number for the Rydberg state. The angular dependence of $J_{z}$, which is responsible for the non-monotonic nature of the interactions, can be explored by varying the angle between the laser propagation direction and the relative vector connecting the two spins \cite{Reinhard, Glaetzle2015}. Typically for the same quantum number, the effective transverse/longitudinal field strengths can be orders of magnitude smaller than the effective Ising interactions. The parameters chosen in this work for the mesonic dynamics are such that the fields are kept constant over distance, and only $J_z$ varies with distance. 

In the semi-classical limit for which $h_{x}\rightarrow0$, one obtains fermionic excitations that are well described by free domain walls. By increasing the transverse field and setting $h_{z}=0$, these domain walls gain kinetic energy. Also, for a nonzero transverse field, the number of domain walls is only quasi-conserved, but for small enough transverse fields $h_{x}\ll J$, one can project the dynamics to subspaces with a conserved number of domain walls \cite{Rutkevich}. When including the long-range character in the Ising interactions, one obtains confinement potentials for otherwise non-interacting free domain walls \cite{liu2019confined}. For this work, we choose parameters such that the effective Ising interaction is attractive. One issue that arises in the context of Rydberg dressing with van der Waals type of interaction is that the asymptotic decay is too short-ranged ($1/r^6$ scaling) for the individual mesons to be stable \cite{kormos2017real}. This can be resolved by either exploiting the long-ranged dipole-dipole interactions of the Rydberg states \cite{Letscher} or by switching on the longitudinal field, $h^z\neq0$ which also ensures the stable formation of individual meson states \cite{kormos2017real,vovrosh2021confinement}. The latter scheme is used in this work. In either case, the short-distance dynamics of the two mesons can still be governed by the non-monotonic effective interactions obtained from Rydberg dressing.
\begin{figure}[t]
	\centering
	\includegraphics[width=\columnwidth]{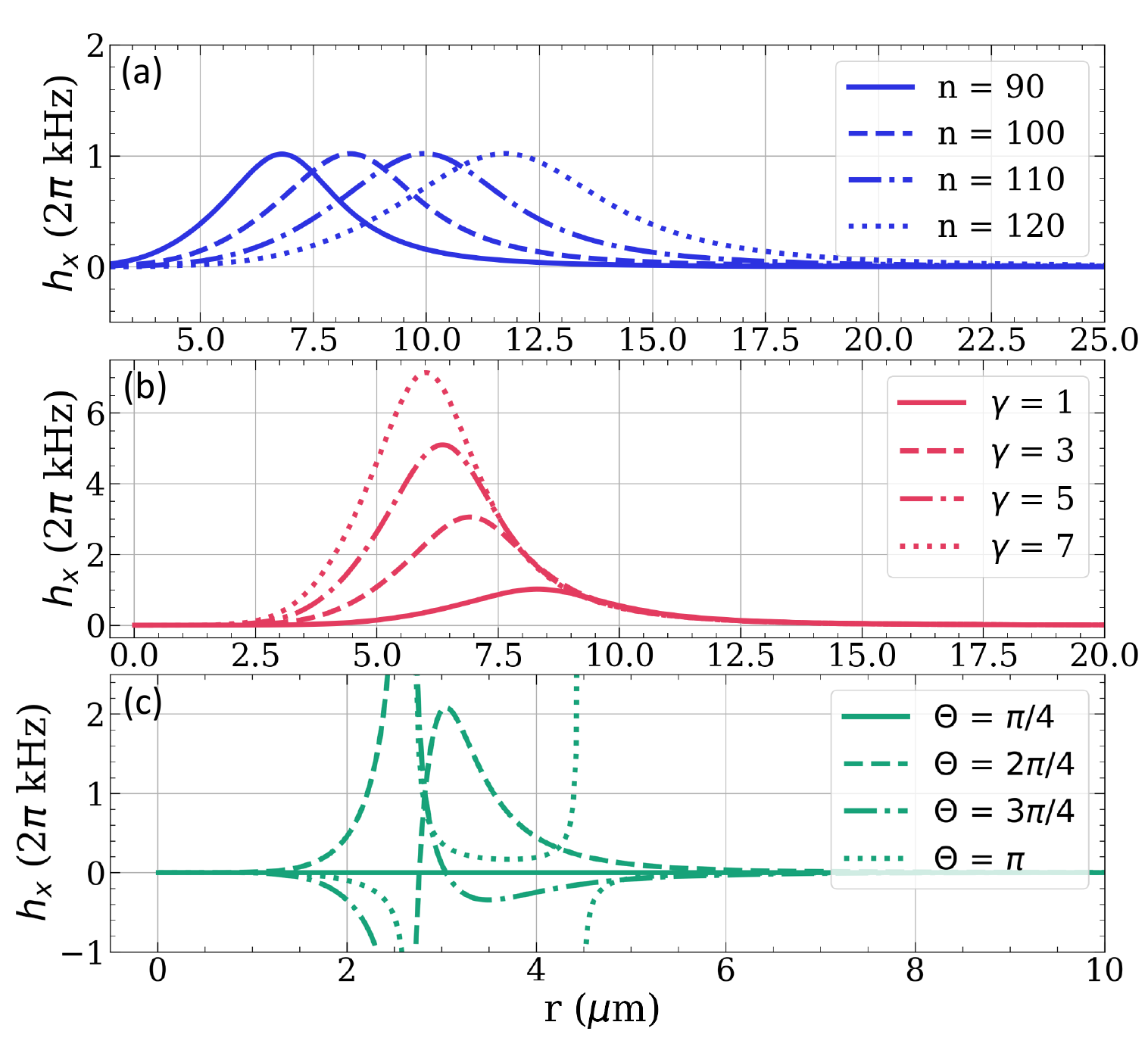}
	\caption{The figure shows the dependence of the transverse field as a function of distance on various parameters. (a) Varying principal quantum number $n$, with fixed $(\Omega_1,\Omega_2,\Delta_1,\Delta_2)= (5, -10, 30, -70)~2\pi$ MHz,  $(\theta, \phi)=(\pi/3, \pi/9)$   (b) Varying the scaling factor $\gamma$, with fixed $n=100$, $(\Omega_1,\Omega_2,\Delta_1,\Delta_2)= (5, -10, 30, -70)\gamma~2\pi $ MHz,  $(\theta, \phi)=(\pi/3, \pi/9)$ (c) Varying $\theta$, with fixed $n=100$, $(\Omega_1,\Omega_2,\Delta_1,\Delta_2)= (5, -10, 30, -70)~2\pi$ MHz,  $\phi=\pi/9$.}
	\label{hx}
\end{figure}

\textit{Mapping to domain-wall physics.\textemdash} The mapping of the Ising model to the mesonic Hamiltonian is well known \cite{kormos2017real, liu2019confined} and more details are provided in Sec.~3-4 of the Appendix. We focus on the simplest case of a pair of 1-mesons (mesons of width $1$ site) located at sites $j_1$ and $j_2$ which is denoted by $\ket{\uparrow \hdots \uparrow\downarrow_{j_1}\uparrow \hdots \uparrow\downarrow_{j_2}\uparrow\hdots \uparrow}$. Using perturbation theory for weak $h^{x}$, the Ising model given in Eq.~\ref{Ising} is mapped to an effective Hamiltonian describing the interaction between the 1-mesons in the reduced basis, which is given as \cite{vovrosh2022dynamical}
\begin{align}\label{tightmeson}
H_{\text{mesons}} =& \sum_{j_1,j_2}h_j [\ket{j_1,j_2+1} + \ket{j_1-1,j_2}]\bra{j_1,j_2}+h.c. \nonumber \\ 
&+ \big(U_j-\frac{4}{j^\alpha}\big)\ket{j_1,j_2}\bra{j_1,j_2} ,
\end{align}	 
where $j=j_2-j_1$, $h_j$ is the inhomogeneous hopping term and $U_j$ is the effective interaction. The hadron-like bound state of two mesons is not coupled to the asymptotic states of free mesons and can have a significant energy gap, especially for small meson sizes. Nevertheless, as already mentioned in the introduction, the transition from free mesons to bound-state mesons can be enhanced either by abruptly changing the transverse field or by modifying the long-range interactions. Both these schemes can be applied to Rydberg setups, although the latter scheme is more challenging to obtain in other cold-atom quantum simulators and is the focus of this work.   Eq.~\ref{tightmeson} is used for the numerical dynamics shown in Fig.~\ref{Fig1_Setup}(b).
\begin{figure}[t!]
	\centering
	\includegraphics[width=\columnwidth]{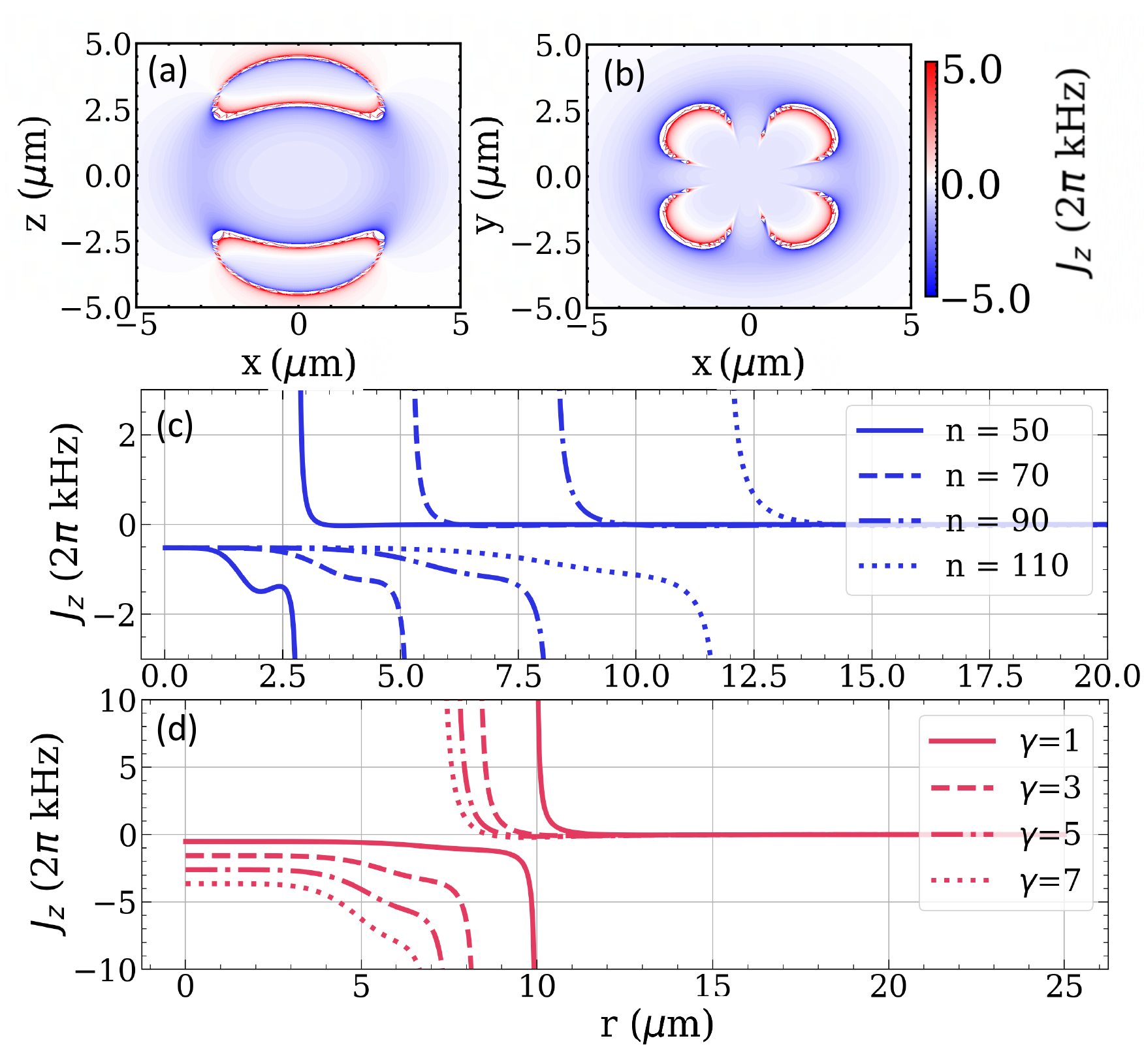}
	\caption{The figure shows the spatial variation of the \textit{resonant-type} effective spin interaction $J^z(\mathbf{r})$ for different parameters. Angular variation of $J^z(\mathbf{r})$ with respect to $\theta$ for $\phi=0$, $n=60$ and $(\Omega_1,\Omega_2,\Delta_1,\Delta_2)= (5, -10, 30, -70)~2\pi$ MHz  in (a) and with respect to $\phi$ for $\theta=\pi/2$ in (b). Radial variation of  $J^z(\mathbf{r})$ for different principal quantum number $n$ for $(\theta, \phi)=(\pi/4, \pi/2)$ and $(\Omega_1,\Omega_2,\Delta_1,\Delta_2)= (5, -10, 30, -70)~2\pi$ MHz in (c) and scaling factor $\gamma$ in $(\Omega_1,\Omega_2,\Delta_1,\Delta_2)= (5, -10, 30, -70) \gamma~2\pi$ MHz for $n=100$ and $(\theta, \phi)=(\pi/4, \pi/2)$ in (d).}
	\label{Fig2_res}
\end{figure}

\textit{Results.\textemdash} Motivated by prior works on quantum simulation of spin models using Rydberg-dressing \cite{Glaetzle2015, Bijnen}, we engineer effective spin interactions for the TFIM that are tailored to achieve successful hadron formation based on \cite{vovrosh2022dynamical}. The key requirement is to have an effective Ising interaction that has a local minimum with a decaying tail in distance. The strength and shape of the effective Ising interaction are controlled by the laser parameters $(\Omega, \Delta)$,  the direction of the excitation laser with respect to the inter-nuclear distance, the lattice spacing, and the principal quantum $n$. In this article, $\text{Rb}^{87}$ atoms were chosen with the two hyperfine sublevels of $\ket{5^2S_{1/2}}$ as the ground states and $\ket{r_1}=\ket{60~^2P_{1/2},m_j = -1/2}$, $\ket{r_2}=\ket{60~^2P_{1/2},m_j = 1/2}$  as the Rydberg states. The van der Waals interactions between Rydberg states for Rb atoms were calculated using a Python package Alkali-Rydberg Calculator (ARC)\cite{Robertson2021}. The anisotropic nature of the dressed potentials, which stems from using non-spherical ($nP$) Rydberg states, is scanned by varying the direction of laser propagation with respect to the lattice axis. The energy scales of the detuning and the van der Waals interactions are considered to be large enough such that the hyperfine interactions in the Rydberg state can be ignored. 

\begin{figure}[t!]
	\centering
	\includegraphics[width=\columnwidth]{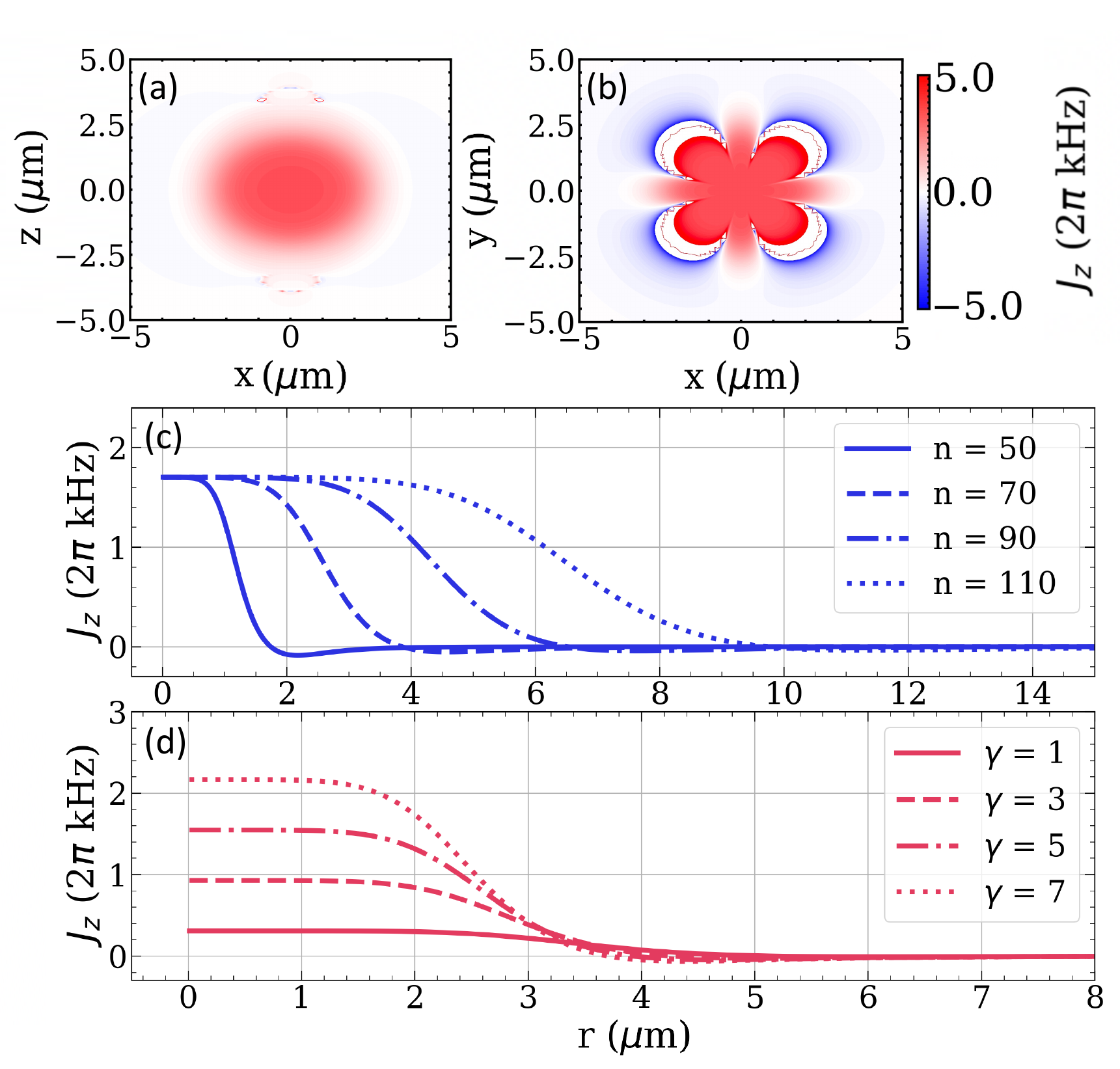}
	\caption{The figure shows the spatial variation of the \textit{step-type} effective spin interaction $J^z(\mathbf{r})$ for different parameters. Angular variation of $J^z(\mathbf{r})$ with respect to $\theta$ for $\phi=0$, $n=60$ and $(\Omega_1,\Omega_2,\Delta_1,\Delta_2)= (0.1, 10, 10, -100)~2\pi$ MHz  in (a) and with respect to $\phi$ for $\theta=\pi/2$ in (b). Radial variation of  $J^z(\mathbf{r})$ for different principal quantum number $n$ for $(\theta, \phi)=(\pi/2, 0)$ and $(\Omega_1,\Omega_2,\Delta_1,\Delta_2)= (0.1, 10, 10, -100)~2\pi$ MHz in (c) and scaling factor $\gamma$ in $(\Omega_1,\Omega_2,\Delta_1,\Delta_2)= (0.01, 1, 1, -10) \gamma~2\pi$ MHz for $n=100$ and $(\theta, \phi)=(\pi/2, 0)$ in (d).}
	\label{Fig3_step}		
\end{figure}

\begin{figure}[t!]
	\centering
	\includegraphics[width=\columnwidth]{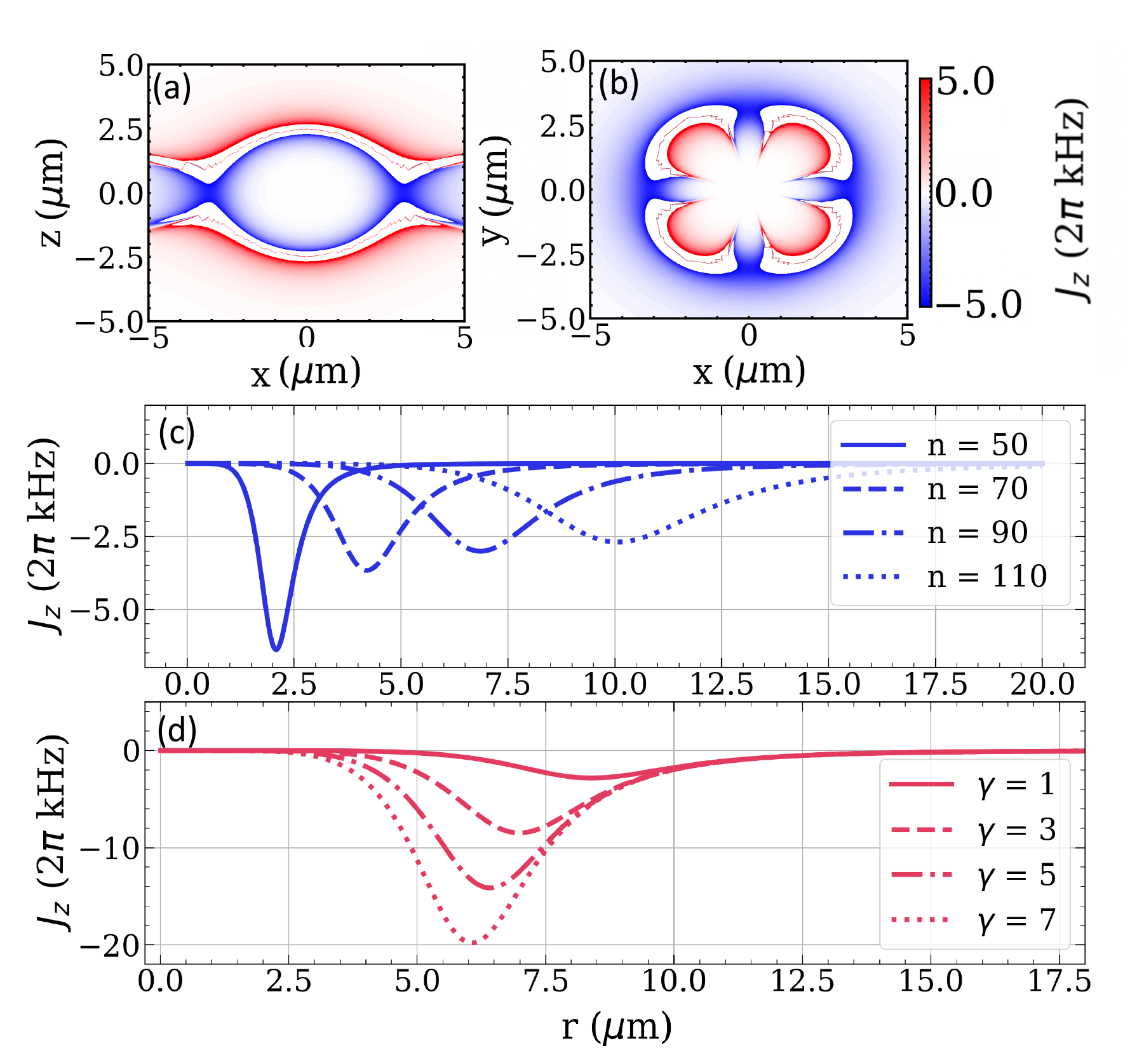}
	\caption{The figure shows the spatial variation of the \textit{hump-type} effective spin interaction $J^z(\mathbf{r})$ for different parameters. Angular variation of $J^z(\mathbf{r})$ with respect to $\theta$ for $\phi=0$, $n=60$ and $(\Omega_1,\Omega_2,\Delta_1,\Delta_2)= (10, 10, 50, -50)~2\pi$ MHz  in (a) and with respect to $\phi$ for $\theta=\pi/2$ in (b). Radial variation of  $J^z(\mathbf{r})$ for different principal quantum number $n$ for $(\theta, \phi)=(\pi/2, 0)$ and $(\Omega_1,\Omega_2,\Delta_1,\Delta_2)= (10, 10, 50, -50)~2\pi$ MHz in (c) and scaling factor $\gamma$ in $(\Omega_1,\Omega_2,\Delta_1,\Delta_2)= (10, 10, 50, -50) \gamma~2\pi$ MHz for $n=100$ and $(\theta, \phi)=(\pi/2, 0)$ in (d).}
	\label{Fig4_hump}
\end{figure}

The dependence of the transverse field on distance for chosen set of parameters is shown in Fig.~\ref{hx}. The typical values of the transverse field is few orders of magnitude smaller than that of the Ising interaction for the same quantum number. Similarly we show the three different types of effective Ising potentials obtained via the Rydberg-dressing of Rb atoms: ``Resonance-type", ``Step-type" and the ``Hump-type" in Figs.~\ref{Fig2_res} - \ref{Fig4_hump}. In these figures, one can see the how the non-monotonic nature of the potentials can be controlled using different parameters.

Figures~\ref{Fig2_res} - \ref{Fig4_hump}(a,b) explicitly depict the much-needed spatial anisotropy in the effective Ising interaction $J^z(\mathbf{r})$. Cases of how the different parameters affect the shape of the interaction are shown in Fig.~\ref{Fig2_res} - \ref{Fig4_hump} (c,d). The position of the resonance, the range of the step function, and the peak of the hump all of which can be varied by simply varying the principal quantum number, while the width of the resonance, the height of the step function, or the hump are controlled by the laser properties. The origin of these shapes can be understood from Eq.~(\ref{u4}). The energies $E_a$ and $E_c$ are fixed in inter-atomic distance and correspond to the energies of the singly excited state in the rotating frame. The distance dependence in energy exists only for $E_b(r)$ which, as explained before, arises from the Rydberg induced Zeeman mixing. For example, the resonance-type potentials in the dressed interaction picture as shown in Fig.~\ref{Fig2_res} occur whenever $E_b(r)$ has a zero crossing in the rotating frame. Since $E_b(r)$ is a non-monotonic function of distance $r$, it is possible to have multiple zero crossings which result in more than one resonance. More non-trivial  Rydberg-dressed potential shapes such as the hump and step (also referred to as soft-core potentials) are also a result of the non-monotonic nature of $E_b(r)$ and are well studied in other works \cite{Henkel2010, Glaetzle2015, Bijnen}.

\begin{figure}[t!]
	\centering
	\includegraphics[width=\columnwidth]{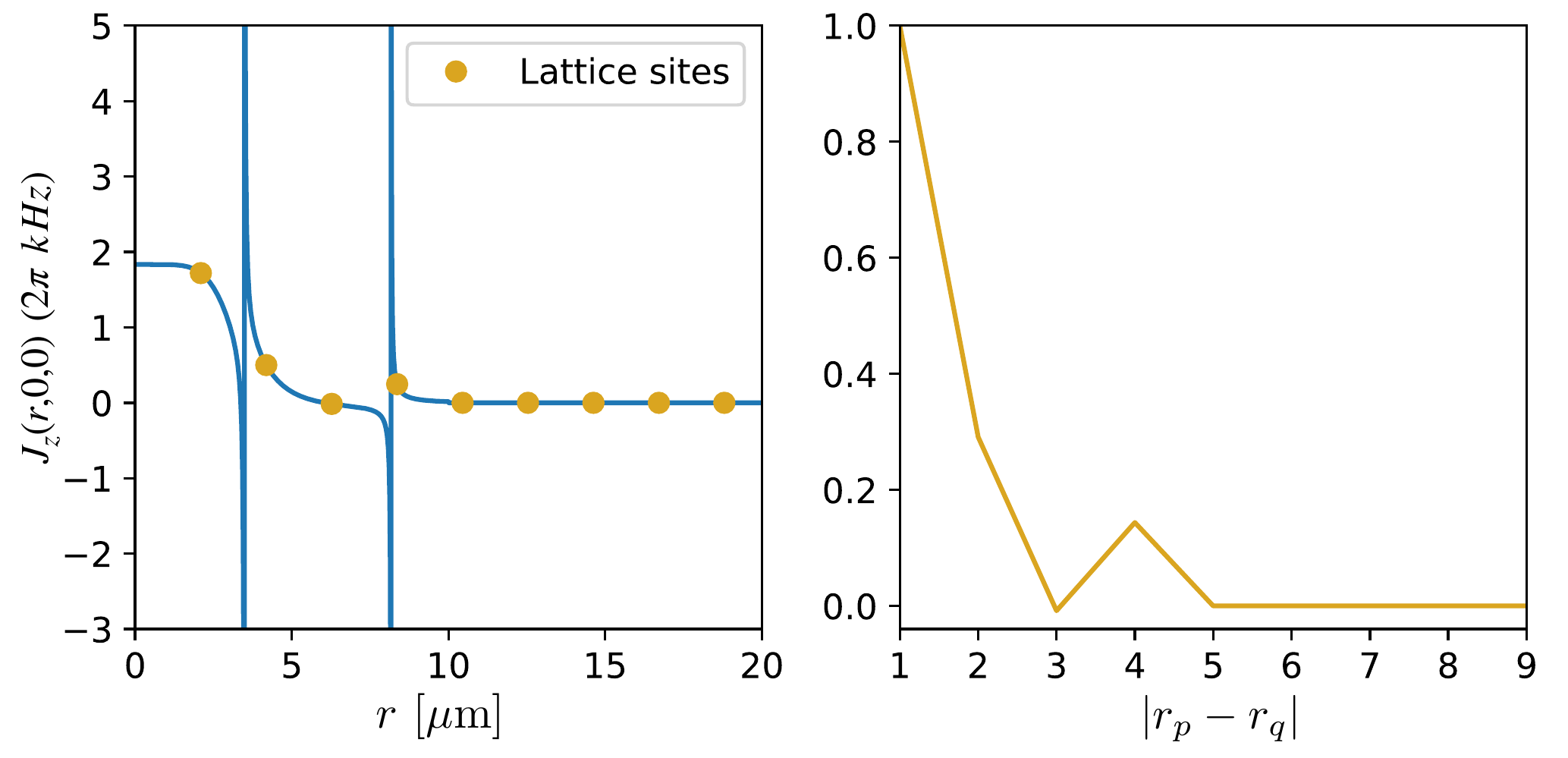}
	\caption{(a) Rydberg-dressed resonance-type potential chosen for the meson dynamics with parameters $n=70$, $(\theta, \phi)=(0, 0)$, $(\Omega_1,\Omega_2,\Delta_1,\Delta_2)= (5, 10, 47, -53)~2\pi$ MHz. (b) By choosing a lattice spacing $a=2.1~\mu$m shown as yellow dots in (a), the effective normalized Ising interaction is obtained for which a tetraquark potential well is produced around site three and is used to obtain the dynamics in Fig.~\ref{Fig1_Setup}(b). Here $|r_p-r_q|$ denote the difference in lattice sites between the pair of mesons.}
	\label{Fig5Ryddressed_pot}
\end{figure}

Both the hump-type potential and the double-resonance type effective potentials are most suitable for hardron formation. However, we take the latter as an example, as shown in Fig.~\ref{Fig5Ryddressed_pot}(a), where a certain fixed lattice spacing is chosen, which is depicted by yellow dots corresponding to the location of the atomic sites. As a result of this choice, one obtains an Ising interaction profile for the dressed atoms as shown in Fig.~\ref{Fig5Ryddressed_pot}(b). The non-vanishing interaction energy at short distances in Fig.~ \ref{Fig5Ryddressed_pot}(b) is needed to provide confinement as was shown in earlier works \cite{kormos2017real,liu2019confined} and we go further than these results by introducing the additional barrier at $|r_p - r_q|\sim$ 4 in order to provide a more stable trap for the tetra-quark after its formation. For future works, it would be interesting to explore the collisions of magnon excitations in the presence of humps (Fig.~\ref{Fig4_hump}) where we anticipate to have very similar results although will lack the analogy with confinement physics. Using the effective Rydberg-dressed potential shown in Fig.~ \ref{Fig5Ryddressed_pot}(b), the dynamics of two $1$-mesons are performed, which is depicted in Fig.~\ref{Fig1_Setup}(b). The individual mesons are initially placed far apart from each other at site number 0 and 30, respectively, where they experience the asymptotic tail of the potential. A non-zero longitudinal field ($|h^z/J^z| = 1$) is also added in order to keep the individual mesons stable. By switching on the transverse field to a value $|h^x/J^z| = 0.3$, the static mesons start moving towards each other. Numerically, the mesonic dynamics is treated by defining a Gaussian wavepacket for each meson which evolves under the Hamiltonian $H_{\text{mesons}}$, and the parameters are provided in Sec.~5 of the Appendix. Thus, two meson wavepackets with opposite momenta approach each other until they see the potential barrier of the effective interaction, which is located at $|r_p - r_q|\sim4$ in Fig.~ \ref{Fig5Ryddressed_pot}(b) or at site 15 in Fig.~\ref{Fig1_Setup}(b). While a large portion of the individual meson wavepacket gets reflected due to the potential barrier, some of it tunnels through the barrier to form a tetraquark. As a result of the quantum tunneling, the lifetime of the meta-stable  bound state can be calculated \cite{vovrosh2022dynamical} which is estimated to be $\tau\sim384.6/J_z\sim 3-15$ ms for $J_z=4-20~2\pi$ kHz. The interaction-induced fusion of mesons occurs at around $t J_z \sim 100$ (see Fig.~\ref{Fig1_Setup}(b)) which is around 795 $\mu$s. The bare Rydberg lifetimes for the states chosen in this work is around 100 $\mu$s which is further enhanced to 10-100 ms due to dressing. This clearly exemplifies the merit of Rydberg-dressing, which not only allows the flexibility to vary the width and location of the double-resonances but also provides an enhanced lifetime due to the fact that the interactions occur effectively between ground state atoms via the weak coupling to Rydberg states.

\textit{Considerations of Experimental feasibility.\textemdash} The timescales provided in the previous section to observe coherent dynamics take into account only the spontaneous emission of the Rydberg state. However, in a real experiment, we have other processes that negatively impact the coherence of the dynamics such as (i) black-body effects, (ii) interaction-induced dephasing that arises from fluctuations in atomic positions, and (iii) the presence of accidental resonances of near-degenerate pair states, especially at small inter-atomic distances which may drastically limit the lifetime of the dressed states. Our setup is envisaged to be in a cryogenic environment ($\sim$10-100 nK) that would suppress the blackbody-radiation-induced transitions to other Rydberg states as well as Doppler broadening similar to most of the current experimental schemes for Rydberg-dressing \cite{Zeiher2016, Steinert, Hollerith}. At few $\mu$m inter-atomic distances, the energy mismatch between two pairs of Rydberg states greatly exceeds their coupling strength resulting in a van der Waals potential. Thus, at such distances, the simplified description of the Rydberg-dressed interactions shown in Figs.~\ref{Fig2_res} - \ref{Fig4_hump} are verifiable to a very good approximation. Naturally, this description becomes increasingly inaccurate for smaller distances which are plagued with a dense region of resonances due to multiple energy crossings from neighbouring pairs of Rydberg states. The density of the spaghetti region experiences a notable decrease beyond a certain distance threshold depending on the principal quantum number. For our case, a distance of around 2 $\mu$m is at the borderline of where effects of the deep spaghetti region qualitatively affect the shape of the potential. The thorough numerical treatment of molecular interaction potentials at such small distances was done in \cite{Bijnen} and is beyond the scope of our work. However, it was confirmed in \cite{Bijnen} and later seen in experiments \cite{Steinert, Hollerith}, that despite the ``spaghetti region of interaction potentials at such small distances", it is possible to obtain potential shapes qualitatively of the type Figs.~\ref{Fig2_res} - \ref{Fig4_hump} provided we choose lattice spacings larger than 1-2 $\mu$m for principal quantum numbers of about $n = 35-70$ for Rb. Interestingly F\"{o}rster-resonant interactions can also be exploited to obtain a more favorable ratio of interactions and decoherence rates which is experimentally achievable. The dynamical behavior of the mesons presented in this work should be observable with current techniques of state-selective imaging of atoms using quantum gas microscopes \cite{Schaus2012, Schauss_2018, Browaeys2020}.

In these setups, it is natural to have the many-body state correspond to the case where all spins are down. The initial state preparation requires a localized spin flip at specific sites to create the four-kink state shown in Fig.~\ref{Fig1_Setup} at $t=0$. This can be achieved using a $\pi$ pulse laser whose width is around few hundred nm. In general, to create a meson of certain size will depend on the lattice spacing and the width of the laser beam \cite{Labuhn2, Bernien}. As mentioned before,  mesons of a certain width with short-range interactions can be stabilized with small but finite longitudinal field. In order to push the individual mesons in opposite directions towards each other, one requires a controlled protocol of spin exchange. For Rydberg dressed-ground states, this can be achieved using protocols similar to \cite{LiLi}. Alternatively, by dressing to excited states one can implement selective spin exchanges via resonant dipole-dipole interaction or by  flip-flop type of van der Waals interaction between Rydberg states \cite{Schempp, Yang}. The issue with microwave dressing to Rydberg states is that one needs to have shorter timescales for coherent dynamics compared to dressed-ground states. The ansatz of describing the wavepacket with certain momenta as a Gaussian needs to be verified experimentally and will depend on various controllable parameters of the hopping term such as the duration and the strength of the microwave dressing laser. Non-Gaussian wavepackets will give similar results depending on the weight of the different velocities that contribute to the wavepacket dynamics but for the benefit of having a better visualization of the formation of hadrons, we chose Gaussian wavepackets that propagate effectively with a single velocity. Moreover, unlike the parameters used in the theoretical simulations, we believe that there is room to allow the collision between mesons with a small uncertainty in momentum space. After the initial kick to the mesons provided by the controlled nearest neighbour spin exchange, the speed of the mesons is related to the transverse field strength $h_x$. However, since our analysis is in the limit of $h_x \ll J$ which guarantees the fixed number of domain wall approximation, we will have slow moving mesons. Although the typical timescales required to observe collisions shown in this work are long compared to what has been achieved in current experiments, we believe that there is scope in optimizing the protocol dynamics, either by minimising the initial meson separation or by maximising $h_x$ while maintaining the domain wall approximation.

\textit{Conclusion and Discussion.\textemdash} Quantum simulation of confinement physics and other high-energy physics phenomena on table-top, controllable experiments is an exciting and growing area of research \cite{Carmen,Surace, Yang_gauge, Zhao_gauge,liu2020realizing}. The formation of long-lived hadronic states by the fusion of individual mesons during their dynamics requires sufficient tunability in the spatial distribution of the interaction profile \cite{vovrosh2022dynamical}. Whilst there exist many ultra-cold platforms, such as trapped ions, dipolar gases, and polar molecules, which have the ability to tune the strength of their long-range interactions, it may be challenging to control the non-monotonic interaction profiles in such setups. On the other hand, Rydberg-dressed interactions have the unique flexibility to reproduce a variety of non-monotonic potentials, some of which were identified in this work in the context of the one-dimensional Ising model. The shape and strength of the effective potential needed to facilitate the creation of hadronic states can be modulated by laser parameters and the choice of the Rydberg state, all of which are controlled in current experiments with sufficient precision. Apart from inheriting the much-needed anisotropy from the Rydberg-Rydberg interactions into the effective spin-spin interactions, the Rydberg-dressed system has an enhanced lifetime compared to the bare Rydberg system making it possible to observe the relevant coherent dynamics leading to hadron formation potentially achievable within experimental timescales. For future work, it would be interesting to see if it is possible to stabilize mesons with larger widths, perhaps by dressing close to F\"{o}rster resonances or with the use of hump-like potentials. There is scope for optimizing the mesonic dynamics such that the time of collision is reduced. Lastly, a more complete study of the dynamics requires taking into account the dissipative and decoherent processes which play a vital role in real experiments.

\begin{acknowledgments}
We are indebted to the fruitful and inspiring discussions with Christian Gross and Rick van Bijnen. JV acknowledges the Samsung Advanced Institute of Technology Global Research Partner- ship and travel support via the Imperial-TUM flagship part- nership. The research is part of the Munich Quantum Valley, which is supported by the Bavarian state government with funds from the Hightech Agenda Bayern Plus.
\end{acknowledgments}

\onecolumngrid
\setcounter{equation}{0}            
\setcounter{section}{0}    
\setcounter{figure}{0}    
\renewcommand\thesection{\arabic{section}}    
\renewcommand\thesubsection{\arabic{subsection}}    
\renewcommand{\thetable}{A\arabic{table}}
\renewcommand{\theequation}{A\arabic{equation}}
\renewcommand{\thefigure}{A\arabic{figure}}
\setcounter{secnumdepth}{2}

\begin{center}
{\large \textbf{APPENDIX}}
\end{center}

\section{Anisotropic Rydberg-Rydberg interaction between two atoms}

A shorthand notation is used for the selected Rydberg states $\ket{nP_{\pm 1/2}m_i;nP_{\pm 1/2}m_j}$ for a pair of atoms which is given as $\ket{m_i,m_j}$. As a result of the van der Waals interaction, the original Rydberg states $\ket{m_i,m_j}$ couple to a bunch of intermediate states $\ket{m_{\alpha}m_{\beta}} \equiv \ket{n_{\alpha},l_{\alpha},j_{\alpha},m_{\alpha};n_{\beta},l_{\beta},j_{\beta},m_{\beta}}$ as permitted by the selection rules. The dipole-dipole interaction operator is defined as
\begin{equation}
	\hat{V}_{\mathrm{dd}}(\mathbf{r})=-\sqrt{\frac{24 \pi}{5}} \frac{1}{r^{3}} \sum_{\mu, \nu} C_{\mu, \nu, \mu+\nu}^{1,1;2} Y_{2}^{\mu+\nu}(\vartheta, \varphi)^{*} \hat{d}_{\mu}^{1} \hat{d}_{\nu}^{2} ,
\end{equation} 
where $\hat{d}_{\mu,\nu \in\{-1,0,1\}}^{i}$  are the spherical basis components of the dipole operator $\hat{d}^{i}$, $Y_{l}^{m}$ are the spherical harmonic functions and  $C_{m_1 m_2; M}^{j_1,j_2;J}$ as the Clebsch-Gordan coefficients. The matrix element $ \bra{m_im_j} \hat{V}_{\mathrm{dd}} \ket{m_{\alpha}m_{\beta}}$  can be expanded into radial and angular parts as follows,
\begin{equation}
	\begin{aligned}
		&\bra{m_im_j} \hat{V}_{\mathrm{dd}} \ket{m_{\alpha}m_{\beta}} = (\bra{n,1,1/2,m_i} \otimes \bra{n,1,1/2,m_j}) \hat{V}_{\mathrm{dd}} (\ket{n_{\alpha},l_{\alpha},j_{\alpha},m_{\alpha}} \otimes \ket{n_{\beta},l_{\beta},j_{\beta},m_{\beta}}) \\
		&= -\sqrt{\frac{24\pi}{5}}\frac{1}{r^3}\sum_{\mu,\nu} C_{\mu, \nu, \mu+\nu}^{1,1 ; 2} Y_2^{\mu+\nu}(\vartheta, \varphi)^* \bra{n,1,1/2,m_i}d_{\mu}^{1}\ket{n_{\alpha},l_{\alpha},j_{\alpha},m_{\alpha}} \bra{n,1,1/2,m_j} d_{\nu}^{2} \ket{n_{\beta},l_{\beta},j_{\beta},m_{\beta}} \\
		&= (-1)^{s-m_{i}} \sqrt{\left(2 \ell_{i}+1\right)\left(2 j_{i}+1\right)\left(2 \ell_{\alpha}+1\right)\left(2 j_{\alpha}+1\right)}\left\{\begin{array}{ccc}
			\ell_{i} & \ell_{\alpha} & 1 \\
			j_{\alpha} & j_{i} & s
		\end{array}\right\}\left(\begin{array}{ccc}
			\ell_{\alpha} & 1 & \ell_{i} \\
			0 & 0 & 0
		\end{array}\right) \\
		& \times(-1)^{s-m_{j}} \sqrt{\left(2 \ell_{j}+1\right)\left(2 j_{j}+1\right)\left(2 \ell_{\beta}+1\right)\left(2 j_{\beta}+1\right)}\left\{\begin{array}{ccc}
			\ell_{j} & \ell_{\beta} & 1 \\
			j_{\beta} & j_{j} & s
		\end{array}\right\}\left(\begin{array}{ccc}
			\ell_{\beta} & 1 & \ell_{j} \\
			0 & 0 & 0
		\end{array}\right) \\
		& \times\left[-\sqrt{\frac{24 \pi}{5}} \sum_{\mu, \nu} C_{\mu, \nu ; \mu+\nu}^{1,1 ; 2}\left(\begin{array}{ccc}
			j_{\alpha} & 1 & j_{i} \\
			m_{\alpha} & \mu & -m_{i}
		\end{array}\right)\left(\begin{array}{ccc}
			j_{\beta} & 1 & j_{j} \\
			m_{\beta} & v & -m_{j}
		\end{array}\right) Y_{2}^{\mu+\nu}(\vartheta, \varphi)^{*}\right]
		\frac{\mathcal{R}^{\alpha}_{i} \mathcal{R}^{\beta}_{j}}{r^3} \\
		&= \mathcal{M}^{ij}_{\alpha\beta} 	\frac{\mathcal{R}^{\alpha}_{i} \mathcal{R}^{\beta}_{j}}{r^3} .
	\end{aligned}
\end{equation}
In the third line, the angular part is expanded using the Wigner-Eckart theorem and is denoted by $ \mathcal{M}^{ij}_{\alpha\beta}$ in the last line. The radial part is expressed as $\mathcal{R}_{\alpha}^{i} = \int r^{2} d r \psi_{n_{i}, \ell_{i}, j_{i}}(r)^{*} r \psi_{n_{\alpha}, \ell_{\alpha}, j_{\alpha}}(r)$. Using the above expressions, the interaction Hamiltonian is written as
\begin{align}
	\hat{H}_{\mathrm{vdw}}(\mathbf{r}) =& \sum_{\alpha,\beta}\frac{ \bra{m_im_j} \hat{V}_{\mathrm{dd}} \ket{m_{\alpha}m_{\beta}} \bra{m_{\alpha}m_{\beta}} \hat{V}_{\mathrm{dd}} \ket{m_{i'}m_{j'}} }{\delta_{\alpha\beta}} \ket{m_im_j}\bra{m_{i'}m_{j'}} \nonumber \\	
	=& \sum_{\substack{\eta = l_{\alpha},l_{\beta} \\ l_{\beta}, j_{\beta}}}  \sum_{n_\alpha, n_\beta}   \frac{\mathcal{R}_i^{\alpha} \mathcal{R}_j^{\beta} \mathcal{R}_{i'}^{\alpha} \mathcal{R}_{j'}^{\beta}}{\delta_{\alpha \beta}r^6} \left[ \sum_{m_{\alpha},m_{\beta}}\mathcal{M}^{ij}_{\alpha\beta} \mathcal{M}_{i'j'}^{\alpha\beta}\right] \ket{m_im_j}\bra{m_{i'}m_{j'}} \nonumber \\		
	=& \sum_\eta  \frac{C_6(\eta)}{r^6} \mathcal{\hat{D}}_{\eta}(\theta,\phi)
\end{align}
where $\delta_{\alpha\beta}= E_{i} + E_{j} - (E_{\alpha} + E_{\beta})$ is the energy difference between the relevant states, $C_6(\eta) =\sum_{n_\alpha, n_\beta} (\mathcal{R}_i^{\alpha} \mathcal{R}_j^{\beta} \mathcal{R}_{i'}^{\alpha} \mathcal{R}_{j'}^{\beta})/\delta_{\alpha \beta}$ is the van der Waals coefficient which determines the strength of the interactions and $\mathcal{\hat{D}}_{\eta}(\theta,\phi) =\sum_{m_{\alpha},m_{\beta}} \mathcal{M}^{ij}_{\alpha\beta} \mathcal{M}_{i'j'}^{\alpha\beta} \ket{m_im_j}\bra{m_{i'}m_{j'}}$ is the angular operator which describes the anisotropy in the interactions. The different channels have been collectively denoted by $\eta$. Based on the properties of the Wigner 3j and 6j symbols, we can deduce the selection rules, which determine that the matrix elements will be non-zero provided the conditions $m_i + \mu = m_{\alpha}$ and $m_j + \nu = m_{\beta}$ are satisfied. Having chosen $\ket{P_{\pm 1/2};P_{\pm 1/2}}$ as our Rydberg states, all the possible inter-mediate states that define the channels are (i) $\ket{S_{1/2};S_{1/2}}$, (ii) $\ket{D_{3/2};D_{3/2}}$ (iii)$\ket{S_{1/2};D_{3/2}}$ and (iv) $\ket{D_{3/2};S_{1/2}}$. 

\section{Effective interactions between Rydberg-dressed ground states for a pair atoms $(\theta,\phi)=(0,0)$}

The unperturbed Hamiltonian for a pair of atoms for $\theta = 0, \phi = 0$ is given as 
\begin{equation}
	\label{0.15}
	\hat{H}_{0}(r)=\left(\begin{array}{cccc}
		V_{22}^{22}(r) - 2\Delta_{1} & 0 & 0 & 0 \\
		0& V_{21}^{21}(r)- \Delta_{1}-\Delta_{2} & V_{12}^{21}(r) & 0 \\
		0 & V_{21}^{12}(r) & V_{12}^{12}(r) - \Delta_{1}-\Delta_{2} & 0 \\
		0& 0& 0 & V_{11}^{11}(r) - 2 \Delta_{2}
	\end{array}\right) ,
\end{equation}
where $V^{ij}_{lk}(r) = \bra{r_ir_j}\hat{H}_{\mathrm{vdw}}(r)\ket{r_lr_k}, \{i,j,l,k\}\in\{1,2\}$ and are explicitly given as
\begin{equation}
	\begin{aligned}
		&V^{11}_{11}(r) = V^{22}_{22}(r) = \frac{2}{81}(2C^{(i)}_{6}+11C^{(ii)}_{6}+14C^{(iii)}_{6})/r^{6} \\
		&V^{12}_{12}(r) = V^{21}_{21}(r) = \frac{2}{81}(4C^{(i)}_{6}+13C^{(ii)}_{6}+10C^{(iii)}_{6})/r^{6} \\
		&V^{12}_{21}(r) = V^{21}_{12}(r) = \frac{8}{81}(C^{(i)}_{6}+C^{(ii)}_{6}-2C^{(iii)}_{6})/r^{6} .
	\end{aligned}
\end{equation} 
The $C^{(i,ii,iii)}_{6}$ are the dispersion coefficients based on the different channels described in the previous section. Diagonalizing the Hamiltonian \(\hat{H}_0\) gives the following eigenstates 
\begin{equation}
	\label{0.17}
	\begin{aligned}
		&\ket{E_1}= \ket{r_2r_2},~\ket{E_2}=\ket{r_1r_1}~~\text{with eigenvalues}~~E_1(r)=V_{11}^{11}(r),~E_2(r)=V_{11}^{11}(r),\\
		&\ket{E_3}=\frac{1}{\sqrt{2}} \left(\ket{r_1r_2}+\ket{r_2r_1}\right) ~~\text{with eigenvalue}~~E_3(r)=V_{12}^{12}(r)+ V_{12}^{21}(r)-\Delta_1-\Delta_2,\\
		&\ket{E_4}=\frac{1}{\sqrt{2}} \left(\ket{r_1r_2}-\ket{r_2r_1}\right) ~~\text{with eigenvalue}~~E_4(r)=V_{12}^{12}(r)- V_{12}^{21}(r)-\Delta_1-\Delta_2.
	\end{aligned}
\end{equation}
As described in the main text, the effective interaction matrix between the two-atom ground states as a result of Rydberg dressing is derived by treating the laser field as a perturbation up to its fourth-order term. In the two atom picture, using the Rydberg-dressed atoms $\ket{\tilde{g}_{i=1,2}\tilde{g}_{j=1,2}}$ as the basis, the effective interactions take the form
\begin{eqnarray}
	\label{equ:2.9}
	&\hat{H}_{eff}(r) &= \left(\begin{array}{cccc}
		\tilde{V}^{22}_{22}(r) & 0 & 0 & 0 \\
		0 & \tilde{V}^{21}_{21}(r) & \tilde{V}^{21}_{12}(r) & 0 \\
		0 & \tilde{V}^{12}_{21}(r) & \tilde{V}^{12}_{12}(r) & 0 \\
		0 & 0 & 0 & \tilde{V}^{11}_{11}(r)
	\end{array}\right) ,\\
	&\text{where}&~~~\tilde{V}^{ij}_{lk}(r) = U^{ij(2)}_{lk} + U^{ij(4a)}_{lk}(r)+U^{ij(4b)}_{lk},~~~\{i,j,l,k\} \in \{1,2\}.
\end{eqnarray}
In the above Hamiltonian, $\tilde{V}^{ij}_{lk} = \tilde{V}^{lk}_{ij}$ due to the symmetry of the Hamiltonian $H$. The matrix elements  are given as
\begin{align}
	&U_{11}^{11(2)}=\frac{\Omega_2^2}{2 \Delta_2},~U_{22}^{22(2)}=\frac{\Omega_1^2}{2 \Delta_1},~U_{21}^{21(2)}=\frac{\Omega_1^2}{4 \Delta_1}+\frac{\Omega_2^2}{4 \Delta_2}, \\
	& U_{11}^{11(4a)}(r) = -\frac{\Omega_{2}^4}{4 \Delta_{2}^2} \frac{V^{11}_{11}(r)-2 \Delta_{1}}{\left(V^{11}_{11}(r)-2 \Delta_{1}\right)\left(V^{11}_{11}(r)-2 \Delta_{2}\right)}, ~U_{22}^{22(4a)}(r) = -\frac{\Omega_{1}^4}{4 \Delta_{1}^2} \frac{V^{11}_{11}(r)-2 \Delta_{2}}{\left(V^{11}_{11}(r)-2 \Delta_{1}\right)\left(V^{11}_{11}(r)-2 \Delta_{2}\right)} \\	
	& U_{12}^{21(4a)}(r) = \frac{\Omega_{1}^2 \Omega_{2}^2}{16 \Delta_{1}^2 \Delta_{2}^2} \frac{\left(\Delta_{1}+\Delta_{2}\right)^2 V^{12}_{21}(r)}{\left(\Delta_{1}+\Delta_{2}-V^{12}_{21}(r)\right)^2-(V^{12}_{21}(r))^2},~U_{21}^{21(4 a)}(r)=\frac{\Omega_1^2 \Omega_2^2}{16 \Delta_1^2 \Delta_2^2}\frac{\left(\Delta_1+\Delta_2\right)^2 \left(\Delta_1+\Delta_2-V_{12}^{12}(r)\right)}{\left(\Delta_1+\Delta_2-V_{12}^{12}(r)\right)^2-\left(V_{21}^{12}(r)\right)^2},\\
	& U_{22}^{22(4 b)} = -\frac{\Omega_1^4}{4 \Delta_1^3},~U_{11}^{11(4 b)} = -\frac{\Omega_2^4}{4 \Delta_2^3},~U_{21}^{21(4 b)} = -\frac{\Omega_1^4}{16 \Delta_1^3}-\frac{\Omega_2^2 \Omega_1^2}{16 \Delta_1^2 \Delta_2}-\frac{\Omega_2^2 \Omega_1^2}{16 \Delta_1 \Delta_2^2}-\frac{\Omega_2^4}{16 \Delta_2^3} .
\end{align}		
where $U_{21}^{21(2)}=U_{12}^{12(2)}$ and $U_{21}^{21(4a,b)}(r) = U_{12}^{12(4a,b)}$. The most general case corresponding to arbitrary angles will have all the off-diagonal terms in the matrix $\hat{H}_{eff} $ to be non-zero. 

\section{Ising model in the basis of the two-domain wall states}

This and the next section have been detailed in  \cite{vovrosh2022dynamical}, which we briefly summarize here. The Hamiltonian for the one-dimensional Ising spin chain  is given as
\begin{equation}\label{Ising2}
	\hat{H} = \sum_{p<q} \left[J_{z}(\mathbf{r}_{pq}) \sigma^z_p \sigma^z_{q} +  h_{x} \sigma^x_p + h_z \sigma^z_p \right] ,
\end{equation}
where all the terms are described in the main text of the article. $h_{x,z}$ have been chosen to be constant over distance. For $h_x=0$, the natural excitations in the model are described by  the non-interacting domain walls. One can now consider the two domain-wall sectors with the basis $\ket{j,n} = \ket{\uparrow...\uparrow\downarrow_j...\downarrow\uparrow_{j+n}...\uparrow}$. This basis $|j, n \rangle$ labels the position of the first domain wall from the left, and $n$ is the relative distance of the second domain wall from the first. On this basis, the Ising model can be re-written as the following,
\begin{equation}
	H = \sum_{j,n}V(n)\ket{j,n}\bra{j,n} - h\big[\ket{j+1,n-1}\\+\ket{j-1,n+1}+\ket{j,n+1}+\ket{j,n-1}\big]\bra{j,n} ,
\end{equation}
and can be understood as a kinetic term for each domain-wall with hopping strength $h$ and a potential $V(n)$. For two domain-walls that are $n$ sites apart feel an attractive potential $V(n)$ of the form \cite{liu2019confined} 
\begin{equation}\label{domainwallpotential}
	V(n) = 4n\zeta(\alpha) - 4\sum_{1\le l<n}\sum_{1\le r < l}\frac{1}{r^\alpha},
\end{equation}
and $\zeta(\alpha)$ is the Riemann zeta function. The most interesting regime is $1<\alpha<2$ where $V(n)$ is unbounded, and pulling the domain walls infinitely far apart  becomes energetically costly, leading to confinement. Nonetheless, for larger $\alpha$, the two-kink subspace shows the presence of deep bound states and asymptotic states of freely propagating domain-walls can be very energetically high with very little overlap with the bound states. However, in order to obtain stable mesons for $\alpha=6$, one requires a non-zero longitudinal field $h_z$.

The study of hadron-like excitations requires considering interactions among mesons. For this purpose, we focus on the two 1-meson subspaces by extending the two domain-wall projections. For the $1-$meson states, we have $n=1$ and can be dropped from the notation. For a pair of  $1-$meson subspace, we use the $|j_1, j_2 \rangle$ notation, which clearly indicates the position of the two individual mesons. On this basis, the projected Hamiltonian takes the shorthand form
\begin{align}\label{fourkink}
	H =& \sum_{j_1, j_2} \big[V(1) \ket{j_1,j_2}\bra{j_1,j_2} + V(2) \ket{j_1,j_2}\bra{j_1,j_2}  + I(j_1,j_2)\ket{j_1,j_2}\bra{j_1,j_2} \big] \nonumber  \\
	&- h \big[ \ket{j_1+1,j_2} + \ket{j_1-1,j_2}+ \ket{j_1,j_2+1} + \ket{j_1,j_2-1} \big] \bra{j_1,j_2} + h.c. 
\end{align}
Here, $I$ can be seen as the meson interaction such that
\begin{equation}
	I(j_1,j_2) = -\frac{4}{(j_1-j_2)^\alpha}
\end{equation}
Note the form of the interaction, $I$, depends on the choice of boundary conditions. Here we consider open boundaries. This effective interaction between mesons falls off as $|j_2 - j_1|^{-\alpha}$ and thus, similar to domain-walls that are spatially distant from each other, individual mesons which are far apart from each other interact only weakly.

\section{Effective Hamiltonian for a pair of 1-mesons}
The mesonic dynamics is constrained to the $1-$meson subspace, which ignores any spin-flip processes (justified for a weak enough transverse field $h$) and is well described with perturbation theory. The Ising model expressed using the two-domain-wall basis states as shown in Eq.~\ref{fourkink} is factored into a diagonal part $H_0$ and an off-diagonal part $V$ proportional to what we will consider the perturbative parameter $h$, that is $H = H_0 + h V$. Using the Schrieffer-Wolff transformation, a unitary transformation is generated by the operator $S$ such that we have the expansion $H^{eff} = e^S H e^{-S} \sim H_0 + h^2 M + O(h^3)$. The Baker-Campbell-Hausdorff expansion of $e^S H e^{-S}$ up to second order in $h$ gives
\begin{equation}
	H^{eff} = e^S H e^{-S} = H_0 + h V + [S, H_0] + h[S, V] + \frac{1}{2}[S, [S, V]] + O(h^3) \ ,
\end{equation}
First-order terms are eliminated by choosing $S$ such that $[S, H_0] = -h V$. The simplest way to do this is element-by-element, enforcing that 
\begin{equation}
	\langle p | S | q \rangle = \frac{ h \langle p| V | q \rangle}{E_p - E_q} \ ,
\end{equation}
where $|p\rangle, |q\rangle$ are eigenvectors of $H_0$ with eigenvalues $E_p, E_q$. With this enforced and absorbing the higher order terms into the $O(h^3)$ factor, the effective Hamiltonian becomes
\begin{equation}
	H^{eff} = H_0 + \frac{h}{2} [S, V] + O(h^3) \ .
\end{equation}
Thus, the resulting effective Hamiltonian contains only on-site potentials and two-site hopping terms. The hopping coefficient $h_{n,n+1}$ determines the moving of $1-$meson by one site. There is also a second order effect in which a $1-$meson can `hop to itself' through a process in which a neighboring spin flips and then flips again to leave the initial state unchanged, $h_{n,n}$ which can be seen as an effective on-site potential. These contributions are explicitly given by 
\begin{equation}
	h_{j} = \frac{h^2}{2}\bigg[\frac{1}{V(1) - V(2) +\frac{4}{{j+1}^\alpha}} + \frac{1}{V(1) - V(2) +\frac{4}{{j}^\alpha}}\bigg],
\end{equation}
\begin{equation}
	U_{j} = 2h^2\bigg[\frac{1}{V(1) - V(2) +\frac{4}{{j+1}^\alpha}}+\frac{1}{V(1) - V(2) +\frac{4}{{j-1}^\alpha}}\bigg],
\end{equation}
where $j=j_2-j_1$ and thus the effective Hamiltonian for a pair of $1-$mesons  can be written as shown in Eq. [9] in the main article. 

\section{Initial meson wavepacket preparation and dynamics}
We theoretically choose Gaussian wavepackets of mesons as initial conditions to allow for high control of meson momentum. In Rydberg atom setups, the many-body product states corresponding to a pair of mesons are readily prepared through local quenches induced by lasers pulses at the appropriate sites. Here, the lattice spacing and the width of the laser beam determine the size of the mesons realized \cite{Omran, Labuhn2,  Bernien}. The momentum at which these mesons propagate along the spin chain can be controlled by the transverse field. The initial states used in the collision of mesons in Fig.~1(b) are given by
\begin{equation}
	\ket{\psi_{t=0}} \propto \sum_{j_1,j_2} e^{-ikj_1-\frac{(j_1+n-2c_1)^2}{4\sigma^2}} e^{ikj_2-\frac{(j_2+n-2c_2)}{4\sigma^2}}\ket{j_1,n,j_2,n},
	\label{wavepacket}
\end{equation}
where $k$ is the initial momentum of each meson, $c_i$ is the center of the wavepacket of the $i^{th}$ meson, and $\sigma$ is the standard deviation of the wavepacket in real space. The parameters for the meson wavepackets in Fig.~1(b) are $q = 1.5$, $\sigma_k = 0.5$ and $\sigma_c = 3$. 

\bibliography{ref}

\end{document}